\documentclass[aps,pra,twocolumn,superscriptaddress]{revtex4-2}

\usepackage[utf8]{inputenc}
\usepackage{amsmath,amssymb}
\usepackage{graphicx}% Include figure files
\usepackage{braket}
\usepackage[colorlinks=true,%
bookmarks=false,%
linkcolor=blue,%
urlcolor=blue,%
citecolor=blue,%
breaklinks]{hyperref}
\usepackage{upgreek,bm}
\usepackage[normalem]{ulem}

\begin{document}

\title{Challenges of variational quantum optimization with measurement shot noise}

\author{Giuseppe Scriva}
\email[]{giuseppe.scriva@unicam.it}
\affiliation{Physics Division, School of Science and Technology, University of Camerino, Via Madonna delle Carceri 9, I-62032 Camerino (MC), Italy}
\affiliation{Institute for Computational Science, University of Zurich, Winterthurerstrasse 190, CH-8057 Zurich, Switzerland}
\affiliation{INFN Sezione di Perugia, Via A. Pascoli, I-06123 Perugia, Italy}

\author{Nikita~Astrakhantsev}
\affiliation{Department of Physics, University of Zurich, Winterthurerstrasse 190, CH-8057 Zurich, Switzerland}

\author{Sebastiano Pilati}
\affiliation{Physics Division, School of Science and Technology, University of Camerino, Via Madonna delle Carceri 9, I-62032 Camerino (MC), Italy}
\affiliation{INFN Sezione di Perugia, Via A. Pascoli, I-06123 Perugia, Italy}

\author{Guglielmo Mazzola}
\affiliation{Institute for Computational Science, University of Zurich, Winterthurerstrasse 190, CH-8057 Zurich, Switzerland}

\date{\today}

\begin{abstract}
Quantum enhanced optimization of classical cost functions is a central theme of quantum computing due to its high potential value in science and technology.
The variational quantum eigensolver (VQE) and the quantum approximate optimization algorithm (QAOA) are popular variational approaches that are considered the most viable solutions in the noisy-intermediate scale quantum (NISQ) era.
Here, we study the scaling of the quantum resources, defined as the required number of circuit repetitions, to reach a fixed success probability as the problem size increases, focusing on the role played by measurement shot noise, which is unavoidable in realistic implementations.
Simple and reproducible problem instances are addressed, namely, the ferromagnetic and disordered Ising chains. 
Our results show that: 
(1) VQE with the standard heuristic Ansatz scales comparably to direct brute-force search when energy-based optimizers are employed. The performance improves at most quadratically using a gradient-based optimizer.
(2) When the parameters are optimized from random guesses, also the scaling of QAOA implies problematically long absolute runtimes for large problem sizes. 
(3) QAOA becomes practical when supplemented with a physically inspired initialization of the parameters.
Our results suggest that hybrid quantum-classical algorithms should possibly avoid a brute force classical outer loop, but focus on smart parameters initialization.
\end{abstract}

\maketitle

\section{Introduction}
Optimization is one of the most anticipated applications of quantum computers due to its commercial value and widespread use in scientific and technological applications~\cite{abbas2023quantum}.
The first argument supporting the benefit of quantum optimization is its ability to search through an exponentially large computational space, of size $N=2^L$, using only $L$ qubits. However, such memory compression alone is not sufficient, as the solution to a classical combinatorial optimization problem is represented by a single (or very few) $L$-bit string. 
This is in contrast with quantum algorithms for solving genuinely quantum mechanics problems, where the source of possible quantum advantage is easier to rationalize~\cite{Feynman1982}.
The quantum computational resource enabling the search is interference. The process begins with a simple, easy-to-prepare quantum state, which undergoes unitary evolution. Ideally, the result of this evolution is such that, when the state is measured, the desired bit string is observed with a high probability~\cite{nielsen_chuang_2010}.

It is still unclear whether quantum optimization offers any advantage over the existing classical methods, such as simulated annealing~\cite{kirkpatrick:1983}.
Interestingly, optimization with quantum annealing has been the first application of commercial quantum devices~\cite{johnson2011quantum,boixo2014evidence}, which mostly rely on incoherent tunneling events to escape the cost-function local minima~\cite{doi:10.1126/science.1252319}. However, it is not easy to prove systematic quantum speedups with analog quantum annealers~\cite{denchev2016computational,albash2018demonstration}, also because quantum Monte Carlo algorithms appear to be able to emulate their tunneling 
dynamics~\cite{isakov2016understanding,mazzola2017quantum,PhysRevA.97.032307,PhysRevB.100.214303}. 
Yet, considerable effort is still ongoing in improving the architecture of these machines~\cite{ozfidan2020demonstration} and their coherence times~\cite{king2022coherent}.

As an alternative quantum optimization strategy, variational quantum algorithms, usually running on digital quantum devices, have gained attention in the quantum computing community due to their  short-depth circuits~\cite{Peruzzo_2014,cerezo2021variational}. 
In this approach, a long quantum state evolution is replaced by a series of short-depth quantum circuits connected through a classical feedback loop.
Variational quantum computation features parametrized circuits that produce a trial state $\ket{\psi({\bm{\theta}})}$.
Its parameters $\bm{\theta}$ are adjusted at every step following an iterative classical procedure. 
The goal is to minimize a cost function $C$, which corresponds to the expectation value $\bra{\psi_{{\bm{\theta}}}} \hat{H}_p \ket{\psi_{{\bm{\theta}}}}$ of the problem Hamiltonian $\hat{H}_p$, or a closely related measure.
At the end of a successful optimization, $\ket{\psi({\bm{\theta}})}$ should be peaked around the solution of the problem.

The two most popular variational algorithms for optimization are the quantum approximate optimization algorithm (QAOA)~\cite{farhi2014quantum} and the variational quantum eigensolver (VQE)~\cite{Peruzzo_2014}. 
Both of them include a parametrized circuit, a classical feedback loop, and a measurement stage. The cost function is evaluated based on the measurement's outcome, and the parameters are adjusted to minimize the cost.

Let us also recall that for combinatorial optimization problems, like Ising spin glasses on general graphs~\cite{barahona:1982}, no polynomial-time algorithm can provably find the global minimum, and the resources to exactly solve these problems scale exponentially with problem size as $\sim 2^{k L}$. 
This is the type of speedup investigated in this article.
While quantum algorithms are not expected to turn the exponential scaling into a polynomial one, the exponent $k$ might be reduced, thus  potentially realizing a substantial speedup over classical algorithms~\cite{doi:10.1126/science.1252319}.

The QAOA method has been the subject of intense studies, including small- and medium-scale hardware experiments~\cite{pagano2020quantum,harrigan2021quantum,pelofske2023quantum,PhysRevE.99.013304}, numerical studies, and theoretical works~\cite{guerreschi2017practical,hadfield2019quantum,guerreschi2019qaoa,zhou2020quantum,moussa2020quantum,boulebnane2021predicting,willsch2020benchmarking,binkowski2023elementary}. Also, VQE optimization has been studied numerically and experimentally~\cite{barkoutsos2020improving,diez2021quantum,amaro2022filtering,zoufal2023variational,kolotouros2022evolving,liu2022layer,Chakrabarti2021thresholdquantum}, and it has been applied to diverse combinatorial optimization problems from protein folding to finance~\cite{robert2021resource,harwood2021formulating,wang2021variational,braine2021quantum}.
However, these previous studies addressed small problem instances, without properly accounting for measurement shot noise. In fact, the latter is unavoidable in physical implementations of practically relevant problem sizes and it might affect the computational complexity of these algorithms. To the best of our knowledge, the scaling of the computational cost for a fixed target success probability, taking into account the measurement overhead to compute the cost function $C$, has not been  exhaustively addressed yet.

The paper is organized as follows. In Sec.\,\ref{sec:models}, we define the testbed problems and the quantum circuits. In Sec.\,\ref{sec:counting}, we introduce the metric to properly assess the computational scaling of the VQE and QAOA algorithms in realistic conditions.
In Sec.\,\ref{sec:vqe}, it is shown that in the presence of quantum measurement noise, VQE displays a scaling not better than the direct space enumeration when energy-based optimizers are used. The situation improves using gradients, computed with the parameters shift rule (see Sec.\,\ref{sec:vqe_grad}), but it remains scaling-wise impractical.
In Sec.\,\ref{sec:qaoa_random}, it is shown that, while showing some scaling improvements, QAOA remains impractical when a full optimization outer loop is required. In this case, the traditional energy-based and a gradient-based optimizer show consistent scalings.
Finally, in Sec.\,\ref{sec:qaoa_init}, we show that QAOA becomes competitive when the parameters are initialized to mimic an adiabatic process. In Sec.\,\ref{sec:conclu}, we draw conclusions and discuss realistic pathways toward quantum advantage in classical optimization problems.

\section{Optimization problems and quantum circuits}
\label{sec:models}
The optimization problems we address correspond to the Ising models defined over $L$ variables $(\sigma_1, \dots, \sigma_L)=\bm{\sigma}$ with $\sigma_j=\pm 1$.
Specifically, we consider the one-dimensional connectivity, nearest-neighbors interactions $J_{j,j+1}$ and local fields $\{h_j\}_{j=1}^L$. The energy of a spin configuration $\bm{\sigma}$ reads
\begin{equation}
E(\bm{\sigma}) = -\sum_{j=1}^{L-1} J_{j,j+1} \sigma_j \sigma_{j+1} - \sum_{j=1}^{L} h_j \sigma_j.
\label{eq:ising}
\end{equation}
Representing a generic spin configuration $\bm{\sigma} \in \{1, -1\}^L$ as a binary string $\bm{x}  \in \{0, 1\}^L$, and writing the energy as $E(\bm{\sigma}) \rightarrow f(\bm{x})$, we write the problem Hamiltonian $\hat{H}_{\mathrm{P}}$ as a diagonal operator
\begin{equation}
    \hat{H}_{\mathrm{P}} = \sum_x f(x) |x\rangle \langle x | ,
\end{equation}
defined by its diagonal matrix elements 
 $f : \{0,1\}^L \rightarrow \mathbb{R}$.
The classical spin variables $\sigma_j$ are promoted to single-qubit Pauli operators $\hat \sigma_j^z$.

\begin{figure*}[hbt]
\includegraphics[width=\textwidth]{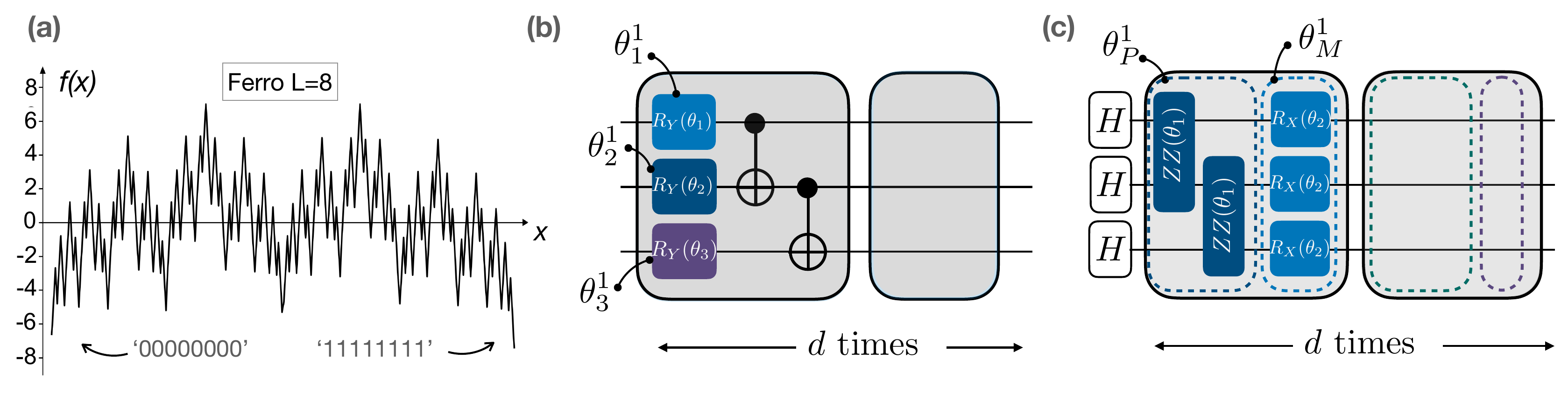}
\caption{\label{fig:circ}
(a) The energy landscape in the computational basis, where items are sorted in the lexicographical order for a ferromagnetic model with $L=8$. The small uniform field breaks the degeneracy between the ``$00\dots0$'' and ``$11\dots1$'' bitstrings, with the latter being the global minimum.
(b) Sketch of RY-CNOT circuits, i.e., a circuit consisting only of $y$ rotations and CNOT gates, used in the VQE and commonly employed in related literature.
(c) Sketch of QAOA circuit featuring the specific problem Hamiltonian.
In both cases we only show the gate decomposition of the first block. The circuit features $d$ blocks, which have the same structure but contain indipendent variational parameters.
}
\end{figure*}

Most analyses reported in this article consider two problem Hamiltonians.
The  first is the \emph{ferromagnetic} Hamiltonian defined by uniform couplings $J_{j,j+1} = J = 1$, and a (small) uniform local field $h_j = h = -0.05$ introduced to break the degeneracy between the two fully-polarized configurations and obtain  a single global minimum.
Despite its simplicity, this model turns out to be hard for most of the considered algorithms.
Its rugged energy surface $f(x)$ is shown in Fig.\,\ref{fig:circ}, where the bitstrings are sorted in the lexicographic order.

The second optimization problem we address is an ensemble of \emph{disordered} Hamiltonians where the couplings and fields are sampled from a normal distribution with zero mean and unit variance:
$J_{j,j+1}, h_j \sim \mathcal{N}(0, 1)$. In this case, 30 realizations of the disorder are simulated for each problem size $L$.

\subsection{VQE with heuristic circuit}
Parametrized quantum circuits~\cite{Peruzzo_2014,cerezo2021variational} are the essential ingredients of any variational quantum algorithm.
These circuits employ parametrized gates, including the single-qubit rotation gates, and multi qubit entangling gates such as the CNOT gate.
The set of  variational parameters $\bm{\theta}$ is optimized in a classical outer loop~\cite{Peruzzo_2014} to minimize a target cost function.

The most commonly studied heuristic circuit is made of $d$ blocks built from a layer of single-qubit rotations $U_\mathrm{R}(\bm{\theta}^l)$ with $l=1,\dots,d+1$ and  an entangling block $U_\mathrm{ent}$ that covers the whole qubit register (see Fig.\,\ref{fig:circ}).
In this article, we consider the entangling block made of a ladder of CNOT gates with linear connectivity, such that the qubit $q_{j-1}$ controls the target qubit $q_j$, and the latter controls the qubit $q_{j+1}$, obeying open boundary conditions.
This choice is commonly used as it mimics the existing sparse qubit connectivity of the quantum hardware.
The layer of single-qubit rotations $U_\mathrm{R}(\bm{\theta}^l)$ acts locally and it corresponds to a tensor product of single-qubit rotations:
\begin{equation}
U_\mathrm{R}(\bm{\theta}^l) = \bigotimes_{j=1}^{L} R_y({\theta}_j^l),
\end{equation}
where $R_y({\theta}^l_j) = \exp{\left( -i \theta^l_j \hat \sigma^y /2 \right)}$ is a rotation around the $y$ axis of the Bloch sphere of the qubit $q_j$, and $l = 1,\dots,d+1$. Here, $\bm{\theta}^l$ denotes an array of $L$ angles.
The full unitary circuit operation is described by
\begin{equation}
U_\mathrm{R-CNOT}(\bm{\theta}) = U_\mathrm{R}(\bm{\theta}^{d+1})~\overbrace{
U_\mathrm{ent} U_\mathrm{R}(\bm{\theta}^{d}) \ldots U_\mathrm{ent} U_\mathrm{R}(\bm{\theta}^{1})}^{\textrm{$d$-times}},
\label{eq:cnotansatz}
\end{equation}
and the final parametrized state reads
\begin{equation}
\ket{\psi(\bm{\theta})} = U_\mathrm{R-CNOT}(\bm{\theta}) \left(\ket{0}^{\otimes L}\right).
\label{eq:trial_ansatz_cnot}
\end{equation}
The total number of variational parameters is $n_{\mathrm{par}} = L(d+1)$.
Notice that we do not use symmetries nor prior knowledge of the optimization problem in building the circuit up.

\subsection{The QAOA circuit}
QAOA can be understood as a digitized version of quantum annealing~\cite{farhi2014quantum} that requires variational optimization of circuit parameters.
These parameters can be seen as the optimizable time steps that control the evolution of the state under the action of the \emph{problem} and the \emph{mixing} operators in a Trotterized fashion.
Notice that the QAOA method precisely dictates the structure of the quantum circuits, while VQE can be implemented with any parametrized quantum circuit. 
In particular, the classical Hamiltonian (i.e., the cost function) explicitly appears in the QAOA circuit, while a VQE circuit may be completely heuristic, with the problem Hamiltonian informing the whole algorithm only through the evaluation of the cost function after the wave function collapses.

The unitary operator defining the Ansatz is made of $d$ blocks, each of them being the product of two unitary operators $\hat{U}_{\mathrm{P}}=\exp{\left( i \theta_{\mathrm{P}}^l \hat{H}_{\mathrm{P}} \right)}$, and
$\hat{U}_{\mathrm{M}}=\exp{\left( i \theta_{\mathrm{M}}^l \hat{H}_{\mathrm{M}} \right)}$, with $l = 1,\dots,d$ and where $\hat{H}_{\mathrm{P}}$ is the problem Hamiltonian, and 
\begin{equation}
    \hat{H}_{\mathrm{M}} = \sum_{j=1}^L \hat{\sigma}_j^x 
\end{equation}
is the nondiagonal \emph{mixing} operator.

The implementation of these unitary operators involves efficient single-qubit rotations along the $x$ axis, denoted as $R_x(\theta) = \exp{\left( i\theta \hat \sigma^x /2 \right)}$, and two-qubit parametrized gates, $R_{zz}(\theta) = \exp{\left( i \theta \hat \sigma^z \otimes \hat \sigma^z /2 \right)}$.
The structure of the QAOA Ansatz implies that all the local $\hat \sigma^z \otimes \hat \sigma^z$ interactions within the same block are ``evolved'' with the same time step $\theta_P^l$, while all the $x$ rotations within the block are parametrized by the same angle $\theta_M^l$ (see Fig.\,\ref{fig:circ}).
The total number of parameters is $n_{\mathrm{par}}=2d$, i.e., is independent of the problem size $L$, and the full unitary operator reads
\begin{equation}
U_\mathrm{QAOA}(\bm{\theta}) = \overbrace{
\hat{U}_{\mathrm{M}}(\theta^{d}_M) \hat{U}_{\mathrm{P}}(\theta^{d}_P) \ldots \hat{U}_{\mathrm{M}}(\theta^{1}_M) \hat{U}_{\mathrm{P}}(\theta^{1}_P)}^{d\textrm{-times}}.
\label{eq:hvansatz}
\end{equation}
The final parametrized state is
\begin{equation}
\ket{\psi(\bm{\theta})} = U_\mathrm{QAOA}(\bm{\theta})~ \left(\frac{\ket{0}+\ket{1}}{\sqrt{2}}\right)^{\otimes L},
\label{eq:trial_ansatz_hv}
\end{equation}
where the initial nonentangled state can be obtained from the  state $\ket{0}^{\otimes L}$ by acting with one Hadamard gate on each qubit.

\section{Resource counting and scaling analysis} 
\label{sec:counting}

\subsection{Statistical noise in evaluating the cost function}
The expectation value of $\hat{H}_{\mathrm{P}}$ over the prepared state is given by the sum of all spin configurations
\begin{equation}
\label{eq:evalop}
    \tilde{C} = \langle \psi_{\bm{\theta}} | \hat{H}_{\mathrm{P}} | \psi_{\bm{\theta}} \rangle = \sum_{x=0}^{2^L-1} |\psi_{\bm{\theta}}(x)|^2 f(x).
\end{equation}
In a realistic setting, the full sum needs to be necessarily approximated using a finite sample of configurations
\begin{equation}
\label{eq:costfunctioneval}
    \tilde{C} \approx \frac1M \sum_{i=1}^{M} f(x_i),
\end{equation}
where $x_i$ are sampled from $|\psi_{\bm{\theta}}(x)|^2$. The precision of this estimate is affected by statistical noise induced by the finite number of quantum measurements $M$.
The error in estimating  $\tilde{C}$ scales as $1/\sqrt{M}$, following the law of large numbers.
We denote this as quantum measurement noise. This noise is very different from hardware noise, produced by qubit's imperfection, as it is rooted in the measurement process of wave functions.
Each quantum measurement requires a circuit repetition.

In numerous studies, Eq.\,\eqref{eq:evalop} is evaluated exactly, which is dubbed the \emph{state-vector} simulation. Instead, in our analysis we account for the effects of the finite $M$.

It has been empirically shown that better performances for optimization problems can be obtained by considering the conditional value at risk (CVaR) estimator of Ref.\,\cite{barkoutsos2020improving}, in which the cost function is evaluated by summing only over the best 25\% of observed outcomes $f(x_i)$:
\begin{equation}
\label{eq:costfunctioneval_cvar}
    {C} = \frac1{M^*} \sum_{i=1}^{M^*} f(x_i).
\end{equation}
Operatively, the $M$ readouts are sorted in nondecreasing order following their output $f(x_i)$, and  only $M^* = M/4$ samples corresponding to the 25\% lowest values are retained.
The value $C$ represents the cost function that is optimized at each iteration. We can also keep track of the current minimum observed value $f_{\mathrm{min}}$, which is generally smaller than $C$. 
Its final value is compared with the exact global minimum of the optimization problem to determine the success rate of the algorithm.
Notice, however, that also when using CVaR one needs to draw $M$ samples.

\subsection{Optimal scaling}
The time complexity of an optimization algorithm can be expressed as the number of function calls $f(x)$ necessary to find the optimum, aiming at a \emph{fixed} success probability as the problem sizes increase.
Each evaluation of the cost function requires $M$ circuit repetitions.

\begin{figure*}[htb]
    \includegraphics[width=\textwidth]{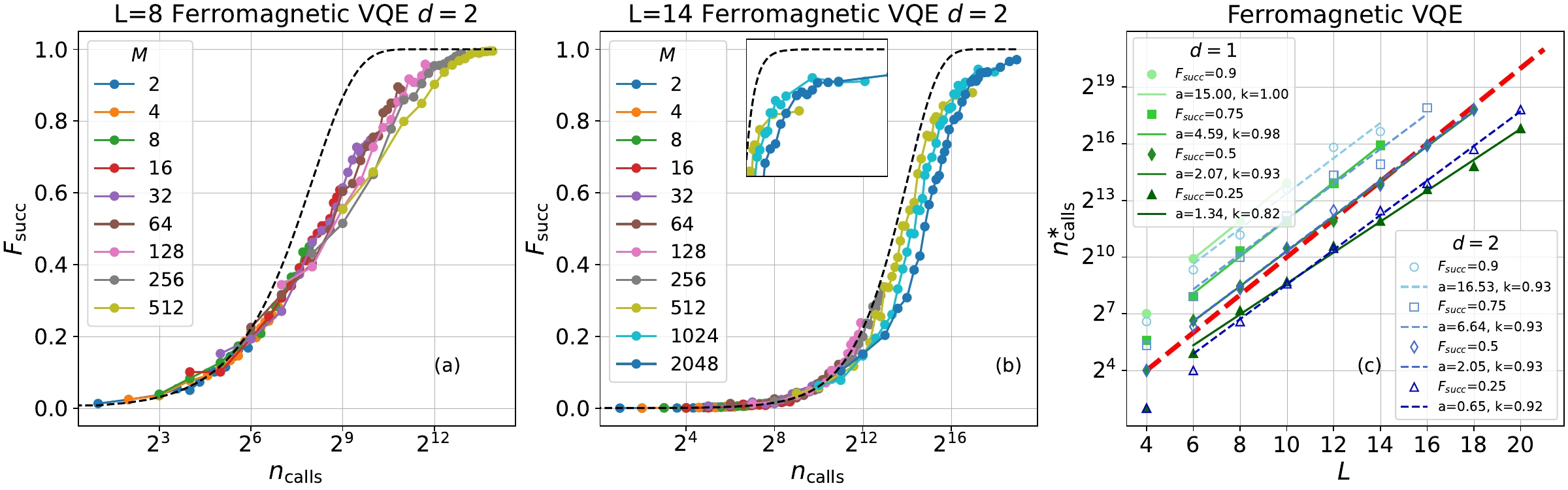}
    \caption{\label{fig:vqe}
    Optimization of the ferromagnetic Ising chain using the VQE Ansatz with depth $d = 2$.
    (a and b): Success probability $F_{\mathrm{succ}}$ as a function of the total number of function calls $n_{\mathrm{calls}}=M\times n_{\mathrm{iter}}$, for (a) $L=8$ spins and (b) $L=14$ spins. Different colors correspond to the different combinations of measurement-shot number $M$ and classical optimization steps $n_{\mathrm{iter}}$. The dashed curve corresponds to the random search with replacement.
    The inset of (b) shows an example of the interplay between $M$ and $n_{\mathrm{iter}}$. To reach larger $F_{\mathrm{succ}}$ it is better to systematically increase $M$. For $F_{\mathrm{succ}} \approx 0.8$, using $M=512$ appears marginally better than $M=1024$, which in turn becomes optimal at $F_{\mathrm{succ}} \approx 0.9$, and so on.
    Crucially, each setup performs worse than the random search.
    (c) Minimal number of function calls $n_{\mathrm{calls}}^*$ as a function of the number of spins $L$ for different $F_{\mathrm{succ}}$. 
    Circuits with $d=1$ (full symbols) and $d=2$ (empty symbols) blocks are considered. The thick dashed (red) line represents the scaling $n_{\mathrm{calls}}^*\sim 2^L$ corresponding to full enumeration.
    Thin continuous and dashed lines represent fitting functions of the form $n_{\mathrm{calls}}^*=a\,2^{kL}$, and the fitting parameters $a$ and $k$, obtained considering the large $L$ regime, are given in the legend.
    All the quantities in this figure and the following are dimensionless. 
    }
\end{figure*}

The total number of function calls required for a full optimization run is therefore
\begin{equation} \label{eq:ncalls}
    n_{\mathrm{calls}} = n_{\mathrm{iter}} \times M,
\end{equation}  
where $n_{\mathrm{iter}}$ is the number of (classical) optimization steps.
The total runtime of the algorithm is proportional to $n_{\mathrm{calls}}$.
A lower bound is given by $t_{\mathrm{run}} = n_{\mathrm{calls}} \times d \times t_{\mathrm{gate}}$, where again $d$ is the circuit depth, expressed as the number of repetitions of a minimal unit (called block) of quantum gates, and $t_{\mathrm{gate}}$ is the time to execute each block.
The value of $t_{\mathrm{gate}}$ strongly depends on the hardware.
In the noisy-intermediate scale quantum (NISQ) era, the gate times can be of order $10$\,ns ($100$\,MHz) for superconducting hardware~\cite{arute:2019}, while digital gate time is predicted to be about $0.1$\,ms ($10$\,kHz) in the fault-tolerant regime~\cite{gidney2019efficient}.
These estimates neglect the qubit reset time, the classical communication, and the measurement time, so they clearly represent optimistic perspectives.

For each problem size, there exists a trade-off between the number of iterations $n_{\mathrm{iter}}$ needed to converge to the global minimum, and the number $M$ of measurements, which controls the accuracy in evaluating the cost function at each step.
Large errors in $C$ may imply slower convergence since the cost function landscape is not correctly reproduced, thus negatively affecting the performance of the classical optimization algorithm.

One of the merits of the present study is the systematic identification of the minimum number of calls, defined as $n_{\mathrm{calls}}^*$, corresponding to the optimal combination of $n_{\mathrm{iter}}$ and $M$ for each problem size $L$, thus enabling a proper scaling analysis.
This concept is similar to the optimal time-to-solution metric developed in quantum annealing~\cite{albash2018demonstration}. We point out that one must have $n_{\mathrm{calls}}^* < 2^L$ to avoid quantum disadvantage~\cite{mazzola2022exponential}, without even discussing the values of $t_{\mathrm{gate}}$.

With the definitions given above, Eq.\,\eqref{eq:ncalls} can be used to compute the number of function calls only in the case of so-called energy-based optimizers. However, in this article, we also consider gradient-based methods (see Secs.~\ref{sec:vqe_grad} and~\ref{sec:qaoa_grad}).
In this case, one needs to compute a $n_\mathrm{par}$-valued array of energy derivatives at each optimization step.
For each parameter, two independent circuit runs need to be executed. This holds both for the parameters shift rule (in this case, when applicable, the gradients are exact) and the finite difference method.
Therefore, the cost for a single iteration has to be computed as $M = 2 n_\mathrm{par} \tilde{M}$, where $\tilde{M}$ is the number of shots per single circuit execution.

\section{Results}

\subsection{Impracticality of VQE}
\label{sec:vqe}
Here we analyze the performance of the VQE method for the ferromagnetic problem. 
The circuit simulations are performed using the open-source Qiskit framework~\cite{Qiskit}.
In evaluating the algorithm's efficiency, a run is considered successful when the absolute minimum is found at least once within the $n_{\mathrm{iter}}$ steps.
This procedure is standard in benchmarking quantum devices, such as quantum annealers, versus classical optimizers~\cite{boixo2014evidence,doi:10.1126/science.1252319,albash2018demonstration}.
The fraction $F_{\mathrm{succ}}$ of successful runs  is estimated considering 1000 executions starting from different (random) initializations of the variational parameters.
It is crucial to note that, within the VQE heuristic circuit, there is no \emph{a priori} method for a smart initialization of the parameters. Therefore, we initialize the parameters using a random uniform distribution of $\bm{\theta}$. Moreover, optimized parameters are not transferable to different instances.

We first inspect how $F_{\mathrm{succ}}$ depends on the total number of function calls $n_{\mathrm{calls}}$, for different problem sizes $L$. 
For each size $L$, several choices of shot numbers $M$ and optimization steps $n_{\mathrm{iter}}$ are considered. Notice that these two parameters determine the number of function calls $n_{\mathrm{calls}}$ [see Eq.\,\eqref{eq:ncalls}]. Importantly, this analysis allows us to identify the minimal number $n_{\mathrm{calls}}^*$, for each target success rate $F_{\mathrm{succ}}$ and for each problem size $L$.
This procedure is crucial to correctly assess the scaling of the computational cost with the problem size. 
Chiefly, it allows us to account for the role of measurement shot noise, which is enhanced for small measurement numbers $M$, while larger $M$ imply a correspondingly larger computational cost for each iteration of the classical optimization algorithm.

In this section, the classical parameter optimization is performed using the constrained optimization by linear approximation (COBYLA) optimizer, a widely adopted energy-based algorithm for QAOA~\cite{barkoutsos2020improving, Weidenfeller2022scaling}.
Let us also recall that the CVaR estimator of Eq.\,\eqref{eq:costfunctioneval} is adopted.
The gradient-based method, which uses the parameters shift rule, is discussed in Sec.\,\ref{sec:vqe_grad}.
The performance of the (COBYLA driven) VQE method, with circuit depths $d=1,\,2$, is shown in  Fig.\,\ref{fig:vqe}.
First, we observe that, for all choices of $M$ and $n_{\mathrm{iter}}$, the value of $n_{\mathrm{calls}}$ required to reach a target $F_{\mathrm{succ}}$ is 
not better than the one corresponding to random search with replacement, see Fig.\,\ref{fig:vqe}(a). 
Furthermore, as shown in Fig.\,\ref{fig:vqe}(c), the minimal number of function calls $n_{\mathrm{calls}}^*$ displays a problematic scaling with the problem size, closely matching the exponential law $n_{\mathrm{calls}}^*\sim 2^{k L}$ with $k\simeq 1$.
This holds for all the thresholds of $0.25 \leqslant F_{\mathrm{succ}} \leqslant 0.9$ considered in this study.
Notably, VQE circuits with depths $d=1$ and $d=2$ display comparable scaling, suggesting that simply increasing the circuit depth does not help.

In Appendix~\ref{appendix:noise} it is shown that hardware noise, which we simulate using a custom model in Qiskit~\cite{Qiskit}, does not significantly affect this scaling.

\subsection{Gradient-based VQE optimization}
\label{sec:vqe_grad}
Next, we consider the VQE algorithm driven by a gradient-based algorithm, addressing again the ferromagnetic problem. 
The gradients are obtained using the parameter shift rule, which is applicable under certain conditions on the adopted gate set~\cite{PhysRevA.98.032309, PhysRevA.99.032331}.
The $n^{\mathrm{th}}$ component of the gradient is computed as $n\in(1,\dots,n_{\mathrm{par}})$:
\begin{equation}
\label{eq:param_shift_rule}
        \frac{\partial \tilde{C}}{\partial \theta_n} = \frac{1}{2} \Big[ \langle \psi_{\bm{\theta}_n^+} | \hat{H}_{\mathrm{P}} | \psi_{\bm{\theta}_n^+} \rangle 
        - \langle \psi_{\bm{\theta}_n^-} | \hat{H}_{\mathrm{P}} | \psi_{\bm{\theta}_n^-} \rangle \Big],
\end{equation}
where $\bm{\theta}_n^\pm=\left(\theta_1,\dots,\theta_n\pm\pi/2,\dots,\theta_{n_{\mathrm{par}}}\right)$.
Notice that, in this case, the cost function is computed as in Eq.\,\eqref{eq:costfunctioneval}, rather than adopting the CVaR estimator.

\begin{figure}[htb]
    \includegraphics[width=0.5\textwidth]{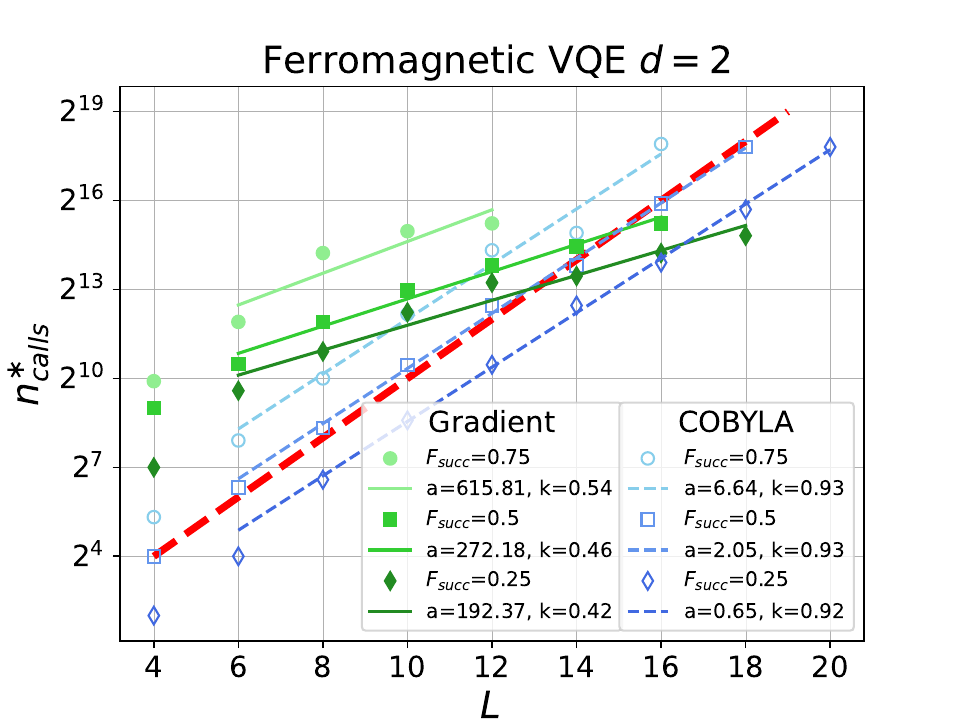}
  \caption{Minimal number of function calls $n_{\mathrm{calls}}^*$ as a function of the number of spins $L$
  for different $F_{\mathrm{succ}}$.
  Thin continuous and dashed lines represent fitting functions of the form $n_{\mathrm{calls}}^*=a\,2^{kL}$, and the fitting parameters $a$ and $k$, obtained by fitting the large $L$ data, are given in the keys.}
    \label{fig:vqe_gradient}
\end{figure}

At each iteration, the parameters $\bm{\theta}$ are updated as
\begin{equation}
    \bm{\theta}^{'} = \bm{\theta} - \eta \nabla{\tilde{C}}\left(\bm{\theta}\right), 
\end{equation}
where $\eta$ is the learning rate. 
The value $\eta=0.1$ is chosen, as it turns out to be reasonably close to optimal from a preliminary analysis on the problem size $L=6$ .
As before, the optimal combination of $M$ and $n_{\mathrm{iter}}$ is found, and the computational complexity is analyzed by observing the scaling of $n_{\mathrm{calls}}^*$ with the problem size (see Fig.\,\ref{fig:vqe_gradient}). Interestingly, an approximately quadratic speedup compared with the COBYLA optimizer is found.

Concluding this subsection, it is worth mentioning that, in the context of quantum chemistry problems, a full quantum eigensolver has been introduced~\cite{doi:10.34133/2020/1486935}. This algorithm implements gradient descent on the quantum device, avoiding the classical optimization step. Future work might focus on adapting this scheme to classical optimization problems.

\subsection{QAOA with random parameters initialization}
\label{sec:qaoa_random}
Here,  the performance of QAOA is analyzed using the (energy-based) COBYLA optimizer. The first tests focus on the ferromagnetic model.
We expect to observe a better performance compared to VQE, because  QAOA features the problem Hamiltonian also in the circuit, not only in the cost function.
To support this intuition, we perform a preliminary comparison, considering circuits with random variational parameters, i.e., avoiding any classical optimization iteration. Specifically, we prepare 1000 different QAOA circuits, and just as many for VQE, using  uniformly distributed parameters, and sample $M=16$ measurements from each of them. The probability $F_{\mathrm{succ}}$ of observing the exact solution at least once is then computed. 
\begin{figure}[htb]
  \includegraphics[width=0.5\textwidth]{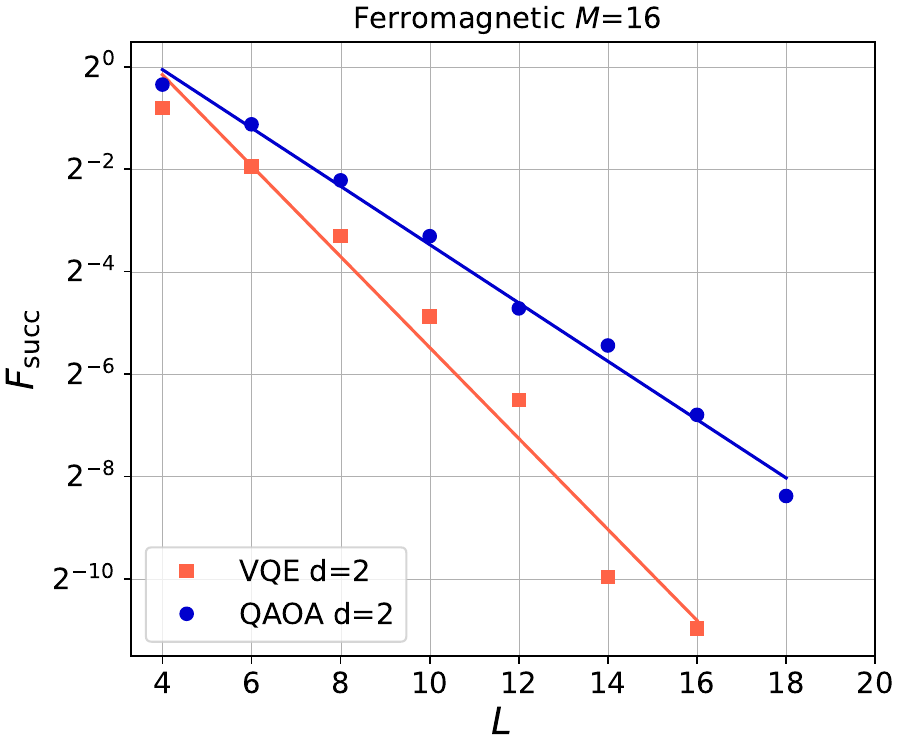}
  \caption{\label{fig:qaoa_vs_vqe_noopt}
  Success probability $F_{\mathrm{succ}}$ as a function of the system size $L$ before the classical optimization, for a fixed shot number $M=16$. The QAOA (circles) and VQE (squares) Ansätze have the same depth $d=2$ and randomly chosen parameters. The lines represent fits, obtained from the large $L$ data, as a guide for the eyes.}
\end{figure}
As shown in Fig.\,\ref{fig:qaoa_vs_vqe_noopt}, the QAOA Ansatz clearly outperforms VQE. We attribute this to its higher degree of localization around the correct solution, even when the parameters are random.
Notice that in this analysis the choice of the optimizer is not relevant, allowing us to compare the circuits independently of the way they are optimized.
One might also expect that, since the QAOA circuit features fewer parameters, it should be easier to optimize as compared to VQE~\cite{PRXQuantum.3.010313}.

\begin{figure*}[htb]
\includegraphics[width=\textwidth]{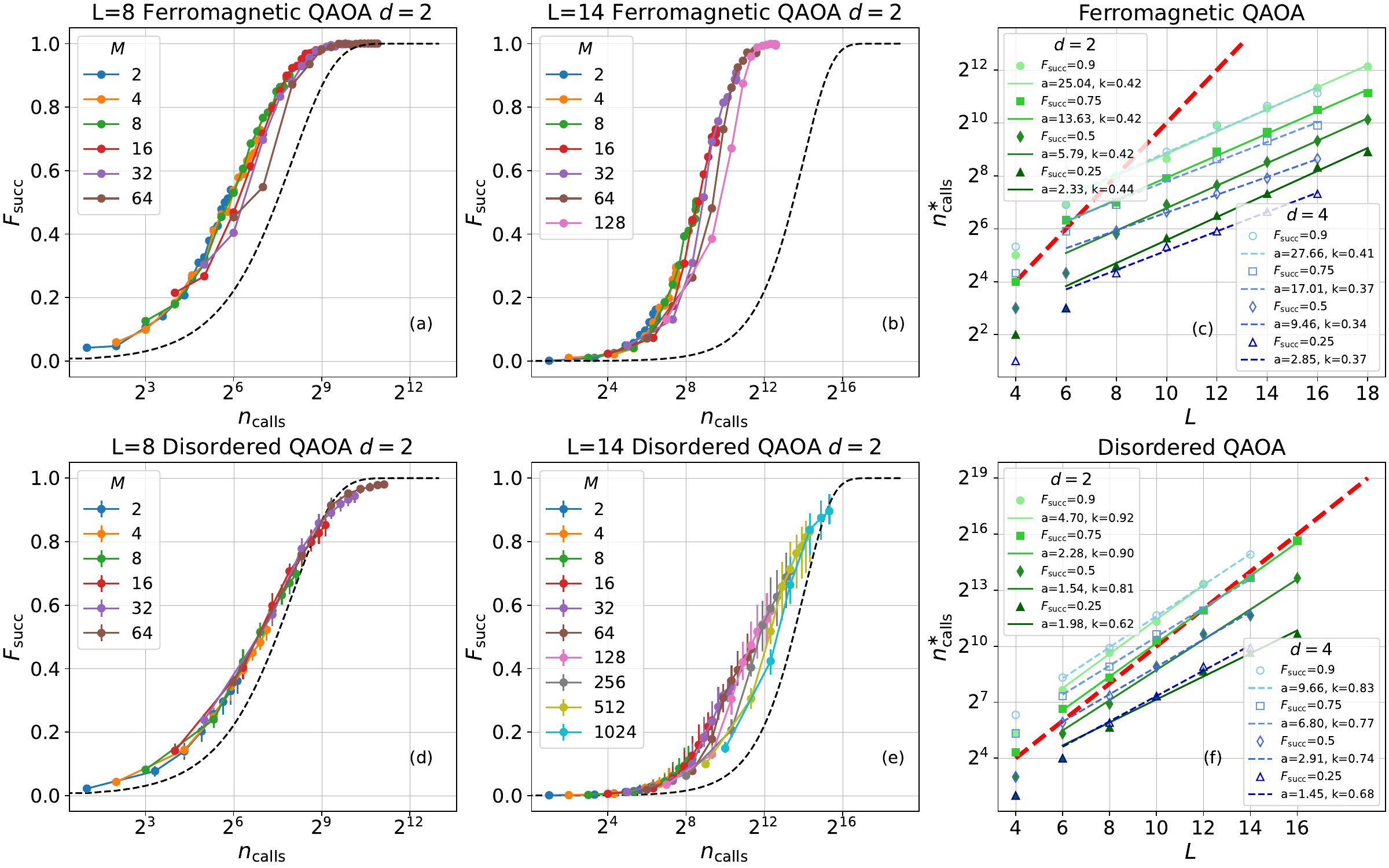}
\caption{\label{fig:qaoa}
    Optimization of the ferromagnetic (first row) and the disordered (second row) Ising chains within the QAOA method.
    (a, b, d, e) Success probability $F_{\mathrm{succ}}$ as a function of the total number of function calls $n_{\mathrm{calls}}$, for (a and d) $L=8$ spins and (b and e) $L=14$ spins. The error bars indicate the 25$^\mathrm{th}$ and the 75$^\mathrm{th}$ percentiles. 
    Different colors correspond to the different combinations of measurement budgets $M$ and classical optimization-step counts $n_{\mathrm{iter}}$. 
    The dashed curve corresponds to the random search with replacement.
    (c, f) The optimal number of function calls, $n_{\mathrm{calls}}^*$, as a function of the number of spins $L$ for different $F_{\mathrm{succ}}$. Circuits with $d=2$ blocks (full symbols) and with $d=4$ blocks (empty symbols) are considered. The thick dashed (red) line represents the scaling $n_{\mathrm{calls}}^*\sim 2^L$ corresponding to exact enumeration.
    Thin continuous and dashed lines represent fitting functions of the form $n_{\mathrm{calls}}^*=a\,2^{kL}$, and the fitting parameters $a$ and $k$ are obtained by fitting the large $L$ data and are given in the keys.}
\end{figure*}

To exhaustively assess the QAOA performance, we repeat the  procedure described in Sec.\,\ref{sec:vqe}, exploring different combinations of $n_{\mathrm{iter}}$ and $M$. Again, this allows us to identify the optimal number of function calls,
$n_{\mathrm{calls}}^*$, for the target cumulative success probability $F_{\mathrm{succ}}$ and problem size $L$.
For each choice of $n_{\mathrm{iter}}$ and $M$, 1000 circuit executions are performed starting from random uniformly distributed parameters.
Notably, for all considered success probabilities $F_{\mathrm{succ}}$, the number of function calls is well described by the exponential scaling law $n_{\mathrm{calls}}^*\sim 2^{kL}$, with $k\simeq 0.4$. This corresponds to an approximately quadratic speedup  as compared to the exact enumeration.
Given that the choice of the classical optimizer may change the observed scaling, in Appendix~\ref{appendix:spsa}, we also test the simultaneous perturbation stochastic approximation (SPSA) algorithm~\cite{705889}. We find that the SPSA and  COBYLA results are compatible.
We further test this finding on a more challenging system, namely, the disordered Ising model.
The results are shown in Figs.\,\ref{fig:qaoa}(d)-\ref{fig:qaoa}(f). Also, in this case they are averaged over 30 realizations of the random couplings and fields of the problem Hamiltonian. 
Similarly to the ferromagnetic case, we observe a profitable scaling, namely, $k\in[0.5,0.8]$, to be compared with the full enumeration, corresponding to $k = 1$. 
However, in this case, extracting the scaling exponent is more difficult, because $n_{\mathrm{calls}}$  needs to be increased to reach large $F_{\mathrm{succ}}$, leading to prohibitive computational times for large problem sizes.
Notably, both for the ferromagnetic and the disordered problem Hamiltonians, increasing the circuit depth from $d= 2$ to 4 does not  substantially affect the scaling.

To summarize the above findings, the observed QAOA scaling exponents are about $k \simeq 0.4$ for the ferromagnetic  problem, and $0.5 \lesssim k \lesssim 0.8$ for the disordered models, using the COBYLA optimizer and random parameters initialization. This scaling is comparable to the one of VQE using gradients. 
While better than full enumeration, these scalings still determine unfeasible runtimes (see discussions in Sec.\,\ref{sec:conclu}) for problem instances of practical interest, i.e., featuring at least  hundreds of spins.

\subsection{Gradient based QAOA optimization}
\label{sec:qaoa_grad}
\begin{figure*}[htb]
    \includegraphics[width=\textwidth]{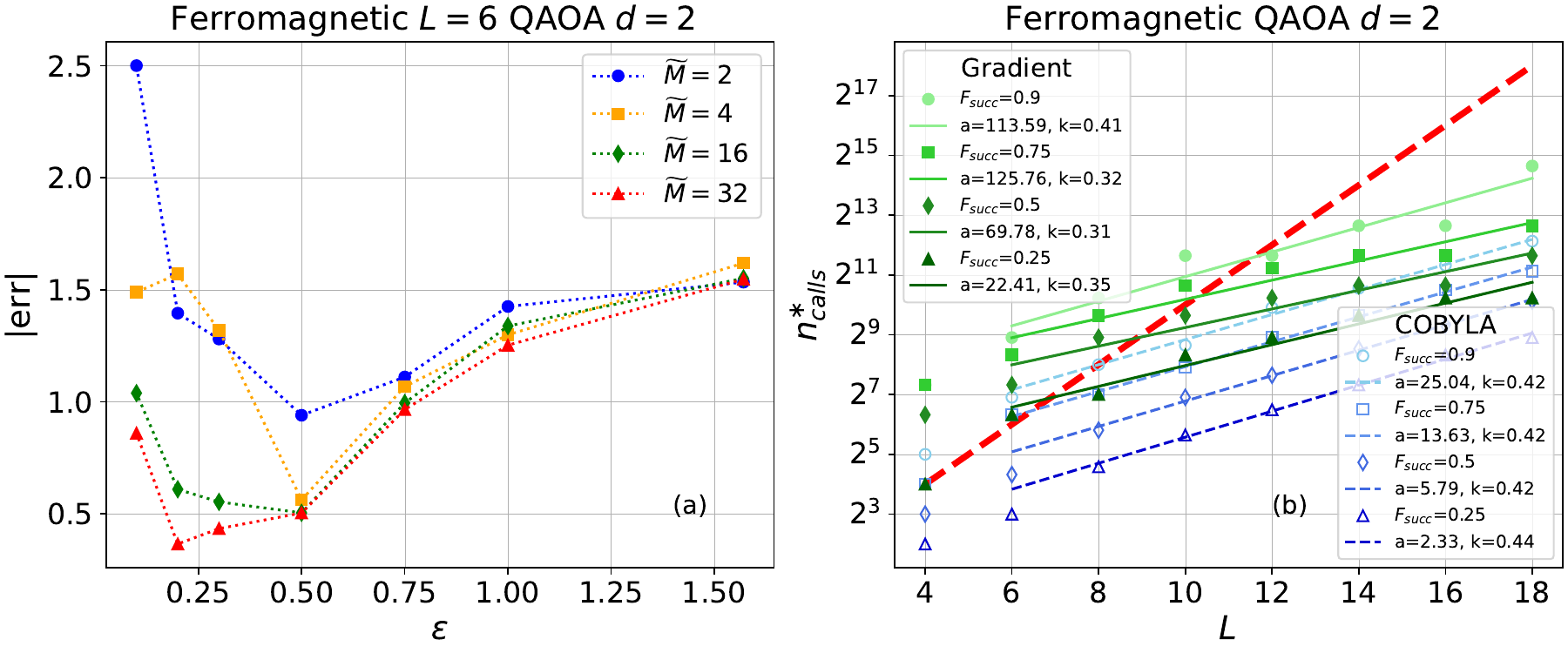}
    \caption{\label{fig:qaoa_gradient}
    (a) Absolute error $|\mathrm{err}|$ in estimating the gradient as a function of the step $\varepsilon$ of the finite difference approximation. We compare results obtained with a different number of shots $M$. 
    (b) The minimal number of function calls $n_{\mathrm{calls}}^*$ as a function of the number of spins $L$, for different success probabilities $F_{\mathrm{succ}}$. Circuits with $d=2$ blocks are considered, with gradient descent (full symbols) and with COBYLA optimizer (empty symbols). The thick dashed (red) line represents the scaling $n_{\mathrm{calls}}^*\sim 2^L$ corresponding to exact enumeration. 
    Thin continuous and dashed lines represent fitting functions of the form $n_{\mathrm{calls}}^*=a\,2^{kL}$, and the fitting parameters $a$ and $k$, obtained considering the large $L$ data, are given in the keys.}
\end{figure*}
Here, we benchmark the scalings of QAOA driven by the COBYLA optimizer against a gradient-based method. Contrary to the case described in Sec.\,\ref{sec:vqe_grad}, the QAOA circuit does not satisfy the assumptions to apply the parameter shift rule~\cite{PhysRevA.99.032331}. 
While there are attempts to extend the parameter shift rule~\cite{wierichs_general_2022}, here we adopt the finite difference approximation. The $n^{\mathrm{th}}$ gradient component is computed as
\begin{equation}
\label{eq:gradfd}
    \frac{\partial \tilde{C}}{\partial \theta_n } = \frac{1}{2\varepsilon} \Big[ \langle \psi_{\bm{\theta}_n^{ +\varepsilon}} | \hat{H}_{\mathrm{P}} | \psi_{\bm{\theta}_n^{ +\varepsilon}} \rangle 
    - \langle \psi_{\bm{\theta}_n^{ -\varepsilon}} | \hat{H}_{\mathrm{P}} | \psi_{\bm{\theta}_n^{ -\varepsilon}} \rangle \Big],
\end{equation}
where $\bm{\theta}_n^{\pm \varepsilon}=\left(\theta_1,\dots,\theta_n\pm \varepsilon,\dots,\theta_{n_{\mathrm{par}}}\right)$ and $\varepsilon>0$ is the  increment.
Small values reduce the finite-difference error, but they also enhance the random fluctuations due to the finite number of measurements $\tilde{M}$ used to estimate the expectation values in Eq.\,\eqref{eq:gradfd}. To identify the optimal trade-off regime, we compare the estimated gradients with the exact results from state-vector simulations [see Fig.\,\ref{fig:qaoa_gradient}(a)]. 
For the typically optimal shot numbers $\tilde{M}\in\left[2,16\right]$, the error is minimized for increments close to $\varepsilon=0.5$. This value is adopted hereafter.

With the above setting, we analyze the scaling of $n_\mathrm{call}^*$ with the problem size $L$ [see Fig.\,\ref{fig:qaoa_gradient}(b)]. Any benefit provided by the gradient turns out to be essentially compensated by its cost in terms of measurement shots. 
Recently, the detrimental cost of gradient estimation has been highlighted addressing the application of quantum computers for electronic structure~\cite{liu2023variational}.
The overall improvement compared to the scaling obtained with the COBYLA optimizer is not sizable. Furthermore, it is worth noticing that even for the larger size considered in this work, the gradient-based optimization requires larger $n_{\mathrm{call}}^*$. 
This is the second main result of the paper: a naive textbook implementation of QAOA using random starting parameters is practically inefficient, even when gradient-based optimizers are used, despite showing an improved scaling with respect to random search.

\subsection{QAOA with annealing inspired parameters initialization}
\label{sec:qaoa_init}
On one hand, the above findings indicate that QAOA is computationally unfeasible for  problem sizes of practical interest. On the other hand, QAOA can be interpreted as a digitized version of quantum annealing, and previous studies have shown that the  available quantum-annealing devices can already find solutions of large-scale spin-glass instances in a reasonable runtime~\cite{doi:10.1126/science.1252319} even for problem sizes as large as $L=512$.
To solve this apparent conundrum, we perform QAOA in its adiabatic limit.
Formally, this limit is reached when $d \rightarrow \infty$.
However, as pointed out in Sec.\,\ref{sec:qaoa_random}, doubling the number of layers does not decisively change  the computational  scaling.
In fact, the number of parameters  increases with the circuit depth, and more parameters usually require more optimization iterations.
\begin{figure*}[htb]
    \includegraphics[width=\textwidth]{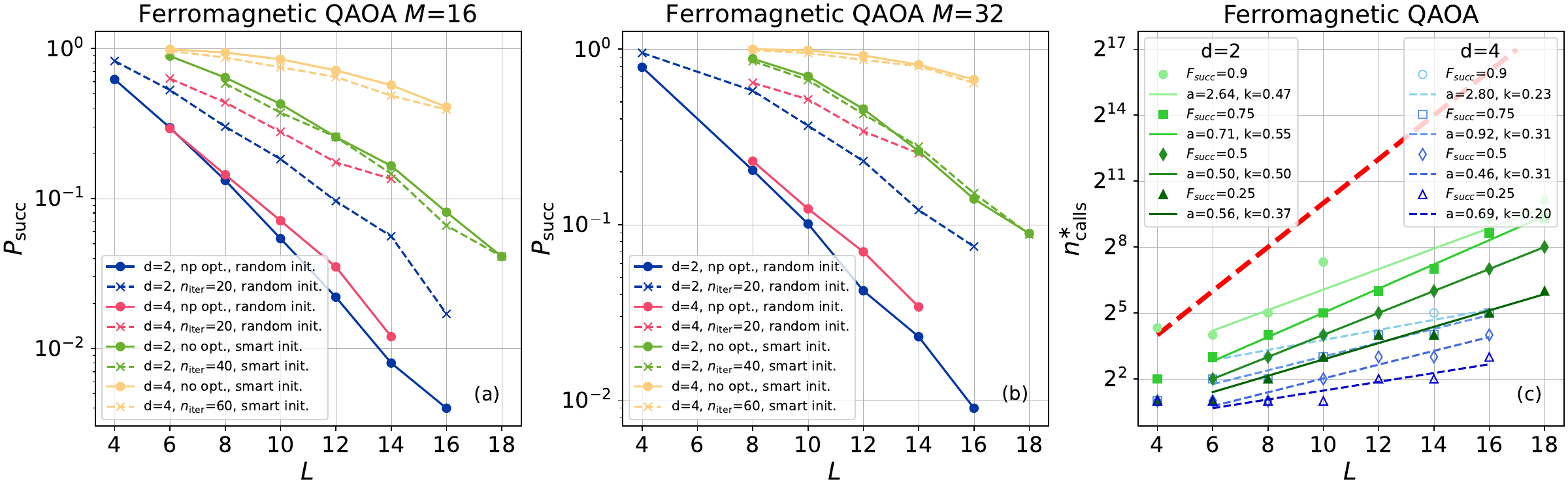}
    \caption{\label{fig:qaoa_smart}
    Comparison of the optimizations starting from random (Sec.\,\ref{sec:qaoa_random}) and annealing-inspired initializations (linear schedule, Sec.\,\ref{sec:qaoa_init}) for the ferromagnetic model. (a and b)  The success probability $P_{\mathrm{succ}}$ as a function of the number of spins $L$, obtained within QAOA before (continuous curves) and after (dashed curves) the classical optimization performed for $n_{\mathrm{iter}}$ steps. The number of shots (a) $M=16$ and (b) $M=32$ are considered, chosen so that $M\ll2^L$. Note that $P_\mathrm{succ}$, i.e., the probability to sample at least once the global minimum at the $n$-th iteration, is equal by definition to $F_\mathrm{succ}$ when $n_{\mathrm{iter}} = 0$. 
    (c) The minimal number of function calls $n_{\mathrm{calls}}^*$ as a function of $L$ at fixed success probabilities $F_{\mathrm{succ}}$, starting from the annealing-inspired initialization. 
    Circuits with $d=2$ blocks (full symbols) and with $d=4$ blocks (empty symbols) are considered. The thick dashed (red) line represents the scaling $n_{\mathrm{calls}}^*\sim 2^L$ corresponding to exact enumeration.
    The thin continuous and dashed lines represent fitting functions of the form $n_{\mathrm{calls}}^*=a\,2^{kL}$, and the fitting parameters $a$ and $k$, obtained by fitting the large $L$ data, are given in the keys.}
\end{figure*}

Still, the analogy with quantum annealing inspires a systematic way to effectively initialize the parameters.
Indeed, in Ref.\,\cite{zhou2020quantum} it was observed that, in state-vector simulations, the optimal parameters often follow a pattern similar to the quantum annealing prescription: the parameters controlling the mixing operator $\bm{\theta}_M$ decrease, while the parameters controlling the problem operator $\bm{\theta}_P$ increase with the layer index.
Following this idea, we initialize the parameters using the simplest discretized linear schedule, as in Ref.\,\cite{sack2021quantum}:
\begin{equation}
\label{eq:smart_init}
    \theta_M^l = \left( 1 -  \frac{l}{d} \right) \Delta_t, \qquad
    \theta_P^l =  \frac{l}{d} \Delta_t,
\end{equation}
where $l \in [1,\dots,d]$.
Notice that in most QAOA literature, the parameters that control the mixing operators are denoted with $\beta$, while the problem parameters are denoted with $\gamma$.

In Ref.\,\cite{sack2021quantum}, this initialization was found effective for MaxCut problems solved via state-vector simulations, i.e., eliminating the measurement shot noise.
Here, we show that this initialization is not only an improvement to the QAOA textbook strategy, but it is also essential to make the algorithm practical in realistic conditions where measurement noise is accounted for.
Notice that with the reparametrization in Eq.\,\eqref{eq:smart_init}, the angles $\theta_M^l$ and $\theta_P^l$  depend only on one real degree of freedom $\Delta_t$.
More complex  reparametrizations could also be possible~\cite{PhysRevA.106.L060401}.

To guide us in the choice of a suitable value for $\Delta_t$, we perform a reasonably exhaustive search, using eight independent repetitions of state-vector simulation using $L=4,6,8,10$ and depths $d=2,4,6,8$.
It is found that the value $\Delta_t \approx 0.80$ is the most frequent outcome of these optimization runs.
Notably, a similar optimal value was found in Ref.\,\cite{sack2021quantum} in the case of MaxCut instances on random graph.
These combined findings suggest that the quantum-annealing-inspired initialization is a general and  robust procedure. This is further corroborated by the results for the disordered Hamiltonian, discussed below.

Hereafter, we first tackle the ferromagnetic problem, using the above prescription. 
The performance of the QAOA circuit with the annealing-inspired parameters is compared to the ones of the QAOA circuits with the parameters obtained after the fixed numbers of optimization iterations $n_{\mathrm{iter}}=20$ and $n_{\mathrm{iter}}=60$, starting from the same smart initialization. 
Notice that here the following definition of success probability $P_\mathrm{succ}$ is adopted: $M$ measurements are performed on the prepared state (with $M=16$ or $M=32$), and the fraction of successful executions at a selected $n_{\mathrm{iter}}$ is recorded. This fraction differs from $F_\mathrm{succ}$, which corresponds to the probability of observing the solution at least once during all optimization iterations, not only in the final state.

The scaling of $P_\mathrm{succ}$ with problem size is shown  in Fig.\,\ref{fig:qaoa_smart}(a) and~\ref{fig:qaoa_smart}(b).

One observes that the QAOA Ansatz with the annealing-inspired linear initialization is already optimal, for both circuit depths $d=2$ and $d=4$.
The optimization of the parameters  does not yield better Ansätze to sample from.
Two important observations are due: (1) the scaling exponent $k$ is reduced compared to the random initialization case, and (2) $k$ decreases with the circuit depth, as opposed to the case of random initialization displayed in Fig.\,\ref{fig:qaoa}(c).
In Fig.\,\ref{fig:qaoa_smart}(c), the scaling of the optimal number of calls, $n_{\mathrm{calls}}^*$, is shown, following the procedure already discussed in Secs.~\ref{sec:vqe} and~\ref{sec:qaoa_random}.
This indicates that optimizations started from random parameters, performed with a finite budget of shots $M$, are not able to showcase the higher expressive power of deeper circuits.
These numerical results suggest that, with a deep enough circuit, the exponent $k$ can be reduced enough to reach practically useful performances for relevant problem sizes. This hypothesis is corroborated by the analysis reported at the end of this subsection.

\begin{figure}[htb]
    \includegraphics[width=0.5\textwidth]{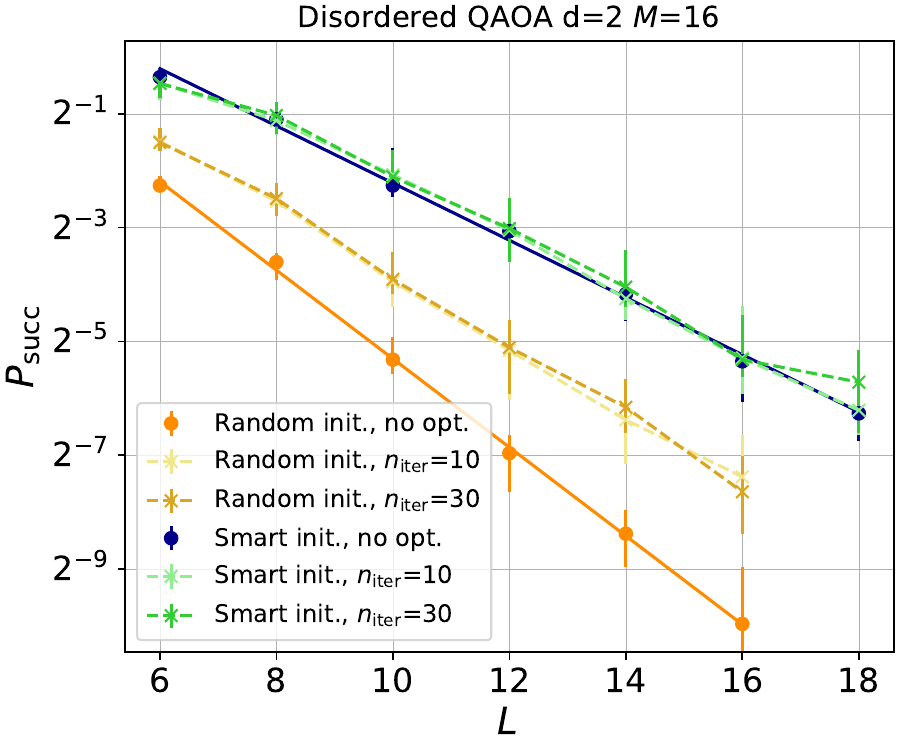}
    \caption{\label{fig:qaoa_smart_disord}
    Success probability $P_\mathrm{succ}$ as a function of the number of spins $L$, starting from random and from annealing-inspired initializations for the disordered Hamiltonian. We compare the success probability $P_\mathrm{succ}$ obtained via QAOA before (continuous curves) and after (dashed curves) the classical optimization performed for $n_\mathrm{iter}$ steps.}
\end{figure}

As anticipated above, here we repeat the numerical experiment using ensembles of disordered Ising chains.
It turns out that the pre-computed value of  $\Delta_t \approx 0.80$ is appropriate, in most instances, also in this setting.
Importantly, in Fig.\,\ref{fig:qaoa_smart_disord} we show that also in this case the smartly initialized Ansatz features almost converged parameters; indeed, the success probability $P_\mathrm{succ}$ does not improve when the circuit is further optimized up to $n_\mathrm{iter}=30$ steps.
Notice that $P_\mathrm{succ}$, i.e., the probability to sample at least once the global minimum at the $n$-th iteration, is equal by definition to $F_\mathrm{succ}$ when $n_{\mathrm{iter}}=0$.
The random-initialized circuit instead benefits from the optimization run, although it never reaches the success probability of the linearly initialized Ansatz.
This result clearly shows that it is much better to use a clever parameters initialization without optimization, instead of randomly initializing the parameters $\bm{\theta}$
 and performing the optimization.

\begin{figure*}[htb]
    \includegraphics[width=\textwidth]{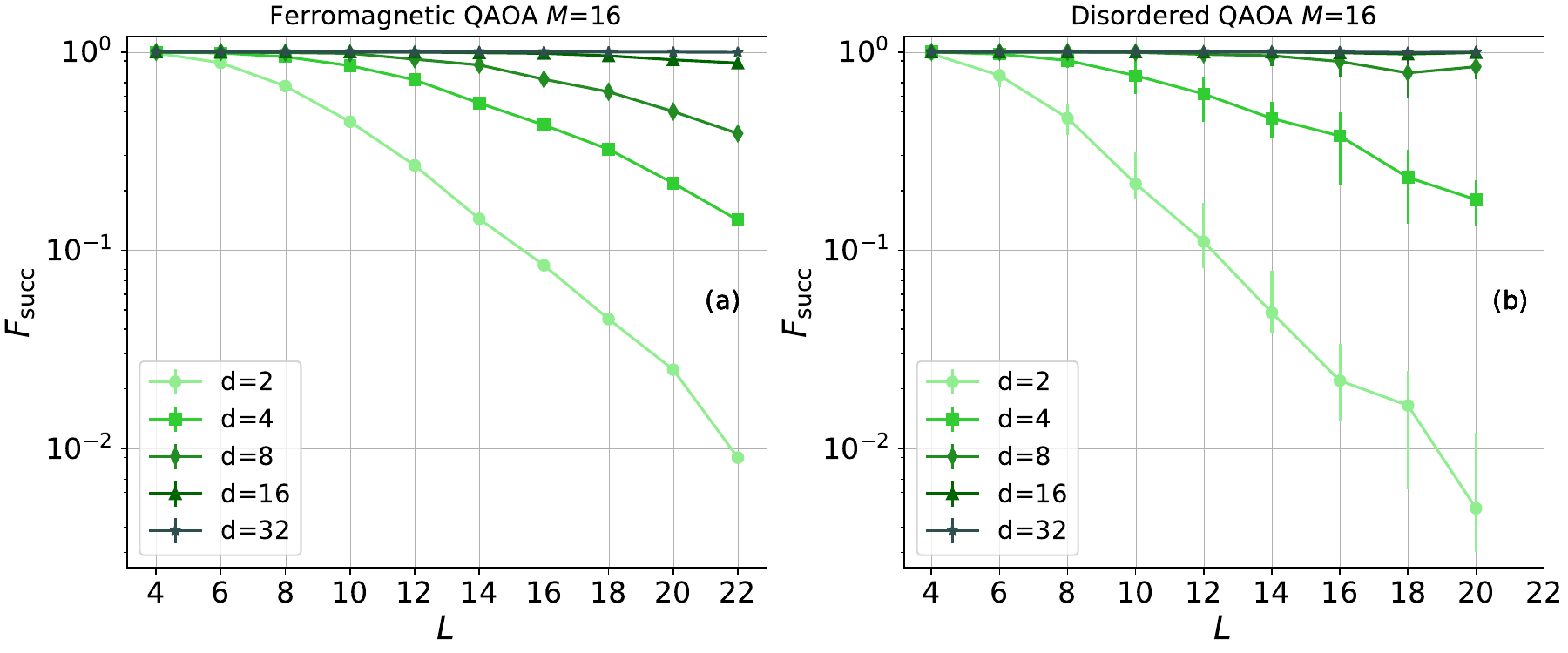}
    \caption{\label{fig:qaoa_smart_vs_depth}
    Success probability $F_{\mathrm{succ}}$ as a function of the number of spins $L$ without classical optimization. Using the smart linear initialization, $F_{\mathrm{succ}}$ grows when the circuit depth is increased, both for the (a) ferromagnetic model and (b) the disordered problems. In the latter, we consider 30 instances of disorder and the error bars indicate the 25$^\mathrm{th}$ and the 75$^\mathrm{th}$ percentiles.}
\end{figure*}

Finally, we try to numerically demonstrate  that, with the annealing-inspired initialization, sufficiently  deep QAOA circuits can reach appealing performances, even without performing classical parameter optimizations.
To this end, we determine the success probability $F_{\mathrm{succ}}$ as a function of the problem size $L$, for several circuit depths $d$ at fixed shot numbers $M$ (see Fig.\,\ref{fig:qaoa_smart_vs_depth}). 
It is found that the performance systematically and rapidly increases with $d$, reaching $F_{\mathrm{succ}}\simeq 1$  even for the largest considered size $L$, for sufficiently deep circuits.
This evidence matches the intuition that QAOA reduces to quantum annealing when $d$ is increased and the circuit parameters follow the pattern in Eq.\,\eqref{eq:smart_init} (although different schedules are possible)~\cite{willsch2020benchmarking}.
Notice that here, the number of shots is $M < 2^L$ for the sizes considered.

\section{Discussion}
\label{sec:conclu}
We critically analyze two popular quantum algorithms for optimization, VQE and QAOA, addressing controllable and reproducible testbed models, i.e., the ferromagnetic and the disordered Ising chains. 
On one hand, our results indicate that, in the practical regime where the number of measurements $M$ per optimization step is much smaller than the Hilbert-space dimension $2^L$, basic  optimization strategies fail to identify suitable circuit parameters.
On the other hand, appealing performances are achieved by deep QAOA circuits when a smart parameters initialization is adopted, as further discussed below.
To reach the above conclusions, we track the total number of measurements $n_{\mathrm{calls}}^*$ to reach a fixed target success probability $F_{\mathrm{succ}}$ in the presence of measurement shot noise, and we analyze its scaling with the problem size $L$.
As expected, we find exponential scalings in the form $n_{\mathrm{calls}}^*\propto 2^{kL}$, and we determine the exponents $k$ considering different setups, including energy-based versus gradient-based classical optimizers in both VQE and QAOA, different circuit depths $d$, as well as random and annealing-inspired parameters initializations in QAOA.

The first result of this article is that VQE shows a very poor scaling with problem size $L$.
When an energy-based optimizer is adopted, the scaling is not better than direct enumeration of the whole computational space, which corresponds to $k=1$.
Introducing additional noise due to simulated hardware errors does not significantly affect the scaling compared to the error-free case.
Notice that our results are not in contrast with existing literature on the use of VQE for classical cost functions~\cite{barkoutsos2020improving,diez2021quantum,amaro2022filtering,zoufal2023variational,robert2021resource}, since these studies report results in the regimes where $M \approx 2^L$ or $n_{\mathrm{calls}} > 2^L$.
We also find that a gradient-based optimization, which we implement in VQE via the parameters shift rule, is useful, leading up to a quadratic speedup compared to the energy-based optimizer COBYLA.

Then we consider QAOA: in contrast to most of the literature, we keep in consideration that the cost function needs to be stochastically evaluated, and in realistic conditions one can afford only $M \ll 2^L$ samples.

We first adopt a textbook version of QAOA, where we optimize the parameters from scratch, i.e., starting from random initial values.
While, as expected,  the total computation complexity is exponential, the exponent $k$ is sizeably reduced compared to the full state enumeration.
Notice that the performance degradation with the system size at fixed quantum resources $n_{\mathrm{calls}}$ is not due here to hardware noise~\cite{harrigan2021quantum}, but to the intrinsic quantum measurement shot noise, a crucial ingredient which is often overlooked and usually leads to overoptimistic expectations for quantum algorithms~\cite{mazzola2022exponential,wecker2015progress}.
Notice also that our numerical findings are compatible with Ref.\,\cite{harrow2021low}, which discusses the query complexity of variational algorithms but only in the vicinity of the global minimum.

With the energy-based optimizer COBYLA, the QAOA scaling exponents turn out to be $k \simeq 0.4$ for the ferromagnetic  problem, and in the range $0.5 < k < 0.8$ for the disordered models; the circuit depth does not significantly affect the scaling.
As opposed to the VQE case, adopting a gradient-based optimization does not sizeably change $k$.
Furthermore, a third optimizer, the SPSA algorithm, provides compatible results.
These scaling exponents can be used to estimate the hypothetical runtimes required to execute the QAOA algorithm on physical quantum devices for realistic problem sizes. 
Assuming the best-observed scenario of the ferromagnetic case, 
$n_{\mathrm{calls}} = 1  \times 2^{0.31 \cdot L}$, some consequential bounds can be provided.
For example, considering the circuit depth $d=2$, gate execution time $t_{\mathrm{gate}}= 10$\,ns for the NISQ era (best case scenario here), one obtains runtimes of about tens of seconds for a hypothetical problem size $L=100$, and a time much beyond the age of universe already  for $L=500$.
These quotes need to be contrasted with tens of milliseconds of total CPU time of simulated annealing~\cite{isakov2015optimised}, or minutes for exact algorithms~\cite{boixo2014evidence} for $L=500$. To achieve a runtime of order 10 ms (resp. minutes), for $L=500$, QAOA should achieve a scaling exponent of about $k=0.04$ (resp. $0.07$).
We conclude that even the best-case scenario observed for the ferromagnetic model is insufficient to provide practical advantage relatively to classical methods or at least feasible absolute times.

Our numerical experiments are consistent with a very recent hardware assessment of QAOA versus quantum annealing, which shows that a $d=2$ QAOA circuit, while  better than random sampling, delivers worse performance than annealing~\cite{pelofske2023quantum}. However, it should be pointed out that, in that large-scale experiment, the performance metric cannot be defined in terms of success probability, since QAOA never provides the exact solution, nor approximate solutions qualitatively comparable with those of simulated or quantum annealing. This is again consistent with our picture.
Moreover, our findings are not in contrast with other previous numerical~\cite{guerreschi2017practical}
or experimental QAOA~\cite{pagano2020quantum} studies, which are either presented in the $n_{\mathrm{calls}} > 2^L$ regime or use a bootstrapping method to initialize the parameters.

To recover an effective algorithm, it is crucial to use a smart initialization of the QAOA parameters.  
In fact, for the simple testbed models we consider, the parameter values given by the annealing-inspired schedule turn out to be very close to the optimal values, such that QAOA provides excellent success probabilities without the need for further parameter optimization.
Interestingly, the same linear schedule proposed in Ref.\,\cite{brandao2018fixed}  for MaxCut problems, based on noise-free simulations, turns out to be suitable also for our ferromagnetic and disordered Ising chains in the presence of measurement shot noise.
While one cannot associate a specific scaling exponent to the smartly initialized QAOA algorithm, as, fortunately, in this case the scaling does improve with the circuit depth, it is quite plausible that sufficiently deep circuits can reach feasible computational times for practically relevant problem sizes.

It is worth  remembering that the proposal to use a smart initialization of the QAOA parameter is  not new~\cite{zhou2020quantum,willsch2020benchmarking,sack2021quantum,PhysRevA.106.L060401}.
However, our findings show that this choice should not only be considered as a good practice to marginally enhance the algorithm efficiency, but it is the only route to make the algorithm practical in the presence of shot noise.
Indeed, if one performs (noise-free) state-vector emulations of the optimization run, good parameters can be recovered anyway, irrespective of the initialization~\cite{zhou2020quantum,willsch2020benchmarking,sack2021quantum}.

Overall, we suggest that future implementation of QAOA should  at least rethink the use of the outer optimization loop, focusing in particular on smart parameter initializations.
While in this manuscript we adopt an annealing-inspired initialization, more flexible solutions, suitable for shallower circuits, are possible.
In general,  the angle array can be re-parametrized as $\bm{\theta} \rightarrow \bm{\theta}(\bm{\alpha})$,  using a smaller number of optimizable parameters $\bm{\alpha}$. This might allow performing fewer optimization steps, similar to the Fourier reparametrization of Ref.\,\cite{zhou2020quantum}.
The research concerning pre-optimization is very active. For instance, the QAOA smart initialization has been studied for the Sherrington-Kirkpatrick model~\cite{Farhi2022quantumapproximate} and the MaxCut problem~\cite{boulebnane2021predicting, 9605328, Egger2021warmstartingquantum}. Moreover, the experimental results achieved in the studies cited above have been analytically confirmed in Ref.\,\cite{PhysRevA.104.L010401}.
Note that the issue of shot noise is well-known in VQE for genuine many-body quantum Hamiltonians, for example, in chemistry~\cite{wecker2015progress}. However, the fact that it also manifests so severely in the case of a classical cost function, which can be measured in a single basis, is important.

We expect our findings to apply in general to variational quantum algorithms strongly relying on a classical optimization loop, but not to other alternatives for quantum-enhanced optimization on digital hardware, including quantum powered sampling~\cite{mazzola:2021,layden2022quantum,mazzola2024quantum}, branch-and-bound algorithm~\cite{PhysRevResearch.2.013056}, and quantum-walks~\cite{callison2019finding}, to name a few proposals.
On a methodological note, these results demonstrate the importance of simple and controllable models to analyze the scaling properties of quantum algorithms in realistic settings.

All data discussed in this article are freely available from Ref.\,\cite{scriva_2023_8223528}.

\begin{acknowledgments}
We acknowledge useful discussion with G.\,Carleo regarding the relation between QAOA and quantum annealing. We thank A.\,Miessen and C.\,Cozza for helping us set up the Qiskit simulation on the cluster.
Computation for the work described in this paper was supported by the Science Cluster at the University of Zurich.
G.S. acknowledges the hospitality of the Institute for Computational Science at the University of Zurich and useful discussions with E.\,Costa.
G.M. acknowledges financial support from the Swiss National Science Foundation (Grant No. PCEFP2{\_}203455).
N.A is funded by the Swiss National Science Foundation, Grant No. PP00P2{\_}176877.
S.P. acknowledges PRACE for awarding access to the Fenix Infrastructure resources at Cineca, which are partially funded by the European Union’s Horizon 2020 research and innovation program through the ICEI project under Grant Agreement No. 800858. 
This work was also supported by the PNRR MUR Project No. PE0000023-NQSTI and by the Italian Ministry of University and Research under the PRIN2022 project ``Hybrid algorithms for quantum simulators'', Project No. 2022H77XB7.
\end{acknowledgments}

\appendix

\section{VQE with hardware errors} 
\label{appendix:noise}

In this Appendix, we inspect the possible role of hardware
errors.
For this, a custom model of hardware noise is introduced, using the open-source Qiskit API~\cite{Qiskit}. A realistic model is obtained, e.g., considering the thermal relaxation due to the qubit environment. Each qubit is then parameterized by a thermal relaxation time constant $T_1=50\,\upmu$s and a dephasing time constant $T_2=70\,\upmu$s.
\begin{figure}[htb]
    \includegraphics[width=0.5\textwidth]{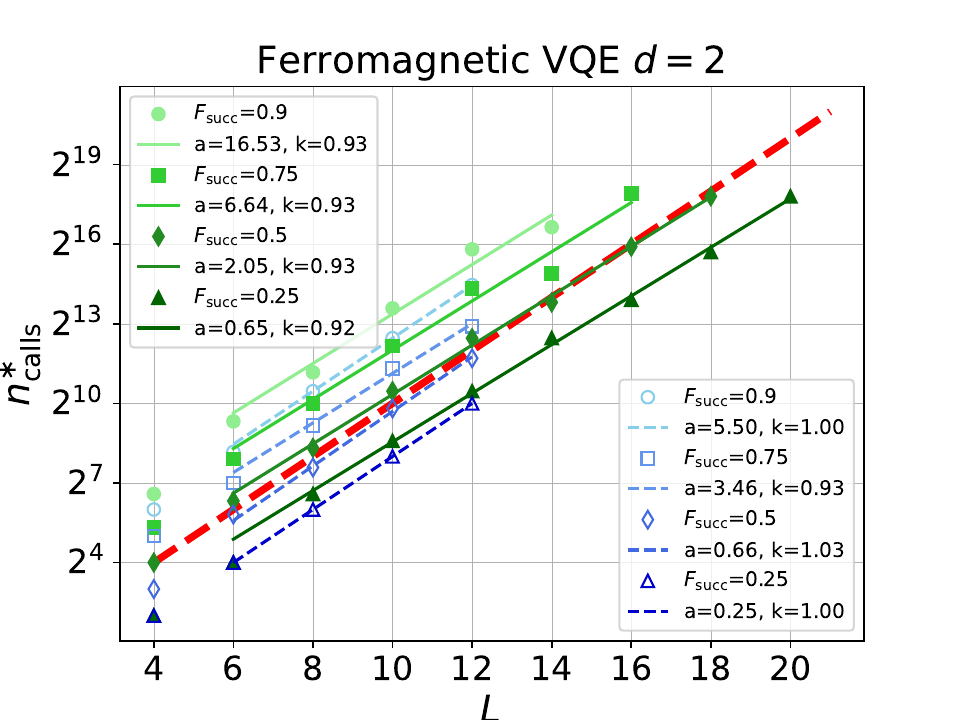}
    \caption{\label{fig:vqe_noise}
    Minimal number of function calls $n_{\mathrm{calls}}^*$ as a function of the problem size $L$, for different $F_{\mathrm{succ}}$. 
    VQE circuits with $d=2$ blocks are considered, both with (empty symbols) and without (full symbols) simulated hardware errors, addressing ferromagnetic chains. The thick dashed (red) line represents the scaling $n_{\mathrm{calls}}^*\sim 2^L$ corresponding to the exact enumeration.
    Thin continuous and dashed lines represent fitting functions of the form $n_{\mathrm{calls}}^*=a\,2^{kL}$, and the fitting parameters $a$ and $k$, obtained considering the large $L$ data, are given in the legend.}
\end{figure}
The performance comparison against the error-free VQE circuits is shown in Fig.\,\ref{fig:vqe_noise}. Ferromagnetic chains are considered, using the CVaR estimator. It turns out that the scaling of $n_{\mathrm{calls}}^*$ with $L$ is not significantly affected by this simulated hardware noise.

\section{QAOA with SPSA}
\label{appendix:spsa}
\begin{figure*}[htb]
    \includegraphics[width=\textwidth]{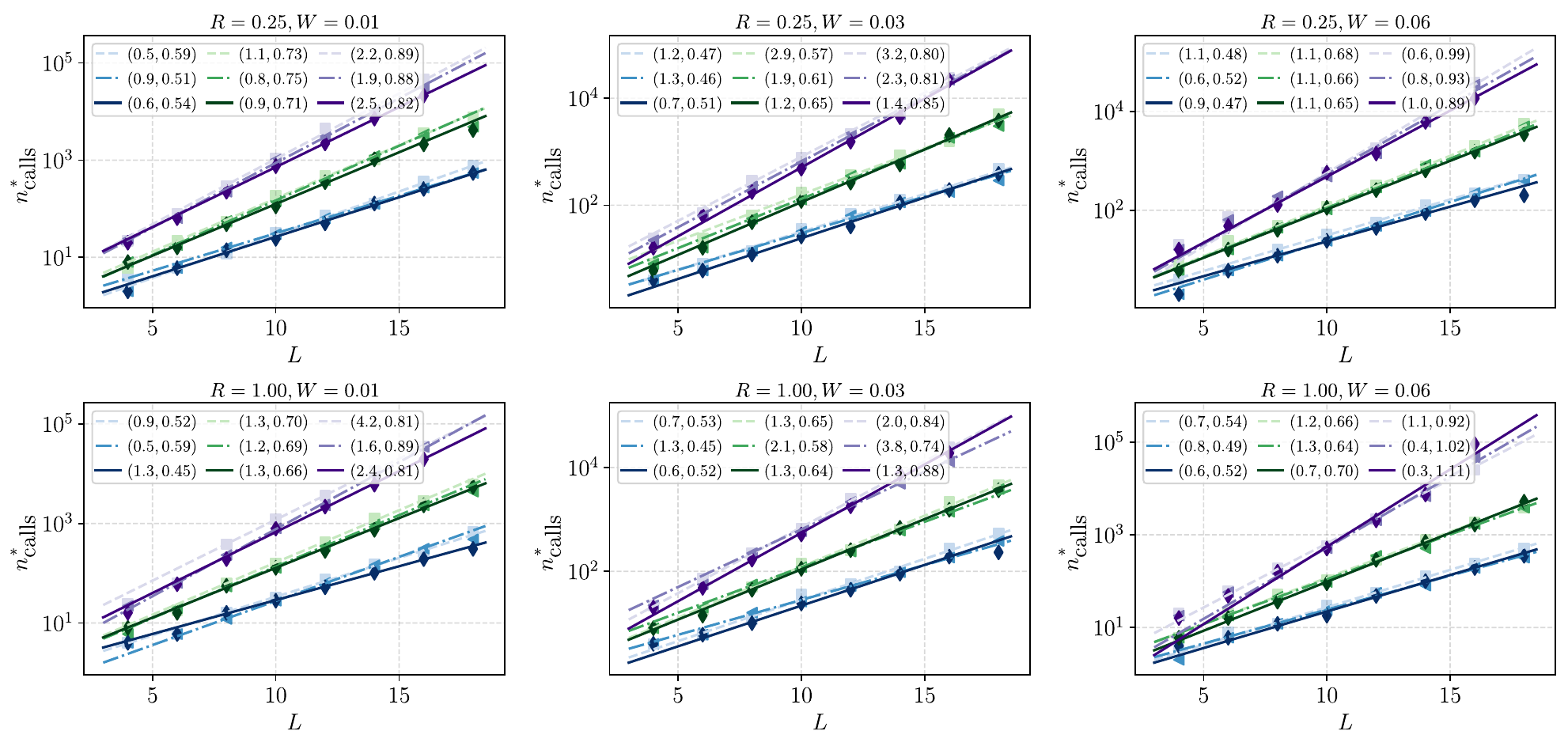}
    \caption{\label{fig:SPSA_panel}
    Scaling of $n_{\mathrm{calls}}^*$ for various depths $d$, CVaR sample fractions $R$ and SPSA proposal lengths $W$. Three curves (from transparent to solid) show $n_{\mathrm{calls}}^*$ for $F_{\mathrm{succ}} = \{0.25,\,0.5, 0.75\}$, respectively. The colors indicate different depths $d = 2$ (blue, lowest data), $d = 3$ (green, intermediate data) and $d = 4$ (purple, highest data). The numbers $(a, k)$ in the brackets in the legends give the optimal values of the exponential fit $n_{\mathrm{calls}}^* = a\, 2^{k L}$ obtained by fitting the data in the regime $L \geqslant 8$.}
\end{figure*}
In this Appendix, we analyze the computational scaling using an energy-based optimizer alternative to COBYLA, namely, the SPSA algorithm~\cite{705889}. This is used to optimize  QAOA circuits of depth in the range $2 \leqslant d \leqslant 4$. The testbed we consider here is the ferromagnetic Ising chain, corresponding to set $J_{j, j+1}=+1$ in Eq.\,\eqref{eq:ising}. 

In the SPSA algorithm, at each optimization step, a random uniformly distributed $n_{\mathrm{par}}$ parameters shift with a constrained length is applied: $\bm{\theta} \to \bm{\theta} + \Delta \bm{\theta}$, with $\|\Delta \bm{\theta}\| \leqslant W$. We consider three values of this maximum norm, namely, $W = \{0.01, 0.03, 0.06\}$. The shift vector is generated as a random vector on a $n_{\mathrm{par}}$-dimensional unit sphere, normalized to length $W$. 
We accept the new parameters if the cost function decreases. As the cost function, we use the CVaR with either 25\% or 100\% of the best-energy samples.

To obtain $n_{\mathrm{calls}}^*$, we use a procedure similar to the one used in Sec.\,\ref{sec:counting}. We consider optimization with $M$ samples generated at each SPSA step, and optimize until $F_{\mathrm{succ}}$ reaches the target value. We then compute $n_{\mathrm{calls}}^* = \min \, \left(M \times n_{\mathrm{iter}} \right).$
The initial parameters are uniform random values in the range $\theta_n \in (-1.0, 1.0)$, and the results for $n_{\mathrm{calls}}^*$ are obtained by averaging over 1000 simulations with random starting points. The results are presented in Fig.\,\ref{fig:SPSA_panel}.

Notably, we observe that the scaling sizeably worsens as $d$ increases from $2$ to $4$. This could be attributed to a more complex optimization landscape which requires a higher $M$ or $n_{\mathrm{calls}}$ to approach the global minimum. At the same time, the Ansatz with $d = 2$ reaches the $k \simeq 0.5$ scaling, therefore it features a quadratic speedup compared with the scaling of the COBYLA-driven VQE optimization, as also found with the QAOA algorithm, driven either by COBYLA or by the gradient-based optimizer.

\bibliography{biblio}

\end{document}